\documentclass[10pt]{iopart}

\usepackage{graphicx}
\usepackage{fancyhdr}
\usepackage{pslatex}   
\usepackage{color}
\usepackage{multirow}

\begin{document}

\title{Shape analysis of strongly-interacting systems: the heavy ion case}



\author{M.A. Lisa$^1$, E. Frodermann$^2$, G. Graef$^{3,4}$, M. Mitrovski$^{5}$,
        E. Mount$^1$, H. Petersen$^6$ and M. Bleicher$^{3,4}$}
\address{$^1$ Department of Physics, Ohio State University, Columbus, Ohio 43210, USA}
\address{$^2$ Department of Physics and Astronomy, University of Minnesota, Minneapolis, Minnesota 55455, USA} 
\address{$^3$ Frankfurt Institute for Advanced Studies, Frankfurt am Main, Germany}
\address{$^4$ Institut f\"{ur} Theoretische Physik, Goethe-Universit\"{a}t, Frankfurt am Main, Germany}
\address{$^5$ Brookhaven National Laboratory, Upton New York 11973, USA}
\address{$^6$ Department of Physics, Duke University, Durham, North Carolina 27710, USA}





\begin{abstract}
Collisions between nuclei at ultrarelativistic energies produce a color-deconfined plasma that
  expands explosively and rapidly reverts to the color-confined (hadronic) state.
In non-central collisions, the zone of hot matter is transversely anisotropic and may be ``tilted''
  relative to the direction of the incoming beams.
As the matter cools and expands into the vacuum, the evolution of the system shape depends sensitively
  on the dynamical response of the plasma under extreme conditions.
Two-pion intensity interferometry performed relative to the impact parameter can be used to measure the
  approximate final shape of the system, when pions decouple from the system.
We use several transport models to illustrate the dependence of the final shape
  on the QCD equation of state and late-stage hadronic rescattering.
The dependence of the final shape on collision energy may reveal non-trivial structures in
  the QCD phase diagram.
Indeed, the few measurements published to date show a tantalizing behaviour in an energy region
  under intense experimental and theoretical scrutiny, as signatures of a first-order phase transition
  may appear there.
We discuss strong parallels between shape studies in heavy ion collisions and those in two other
  strongly-coupled systems.
\end{abstract}

\maketitle

\section{INTRODUCTION}

Color confinement is the most unique and important feature of Quantum Chromodynamics (QCD), believed to be the correct
  field theory of the Strong Interaction.
While the symmetries and much of the dynamics of the interaction are understood, a complete understanding of the confinement
  mechanism is hindered by the difficulty of theoretical calculations in a non-perturbative regime.
Ultra-relativistic heavy ion collisions serve as an ideal, if fleeting, laboratory for the study of deconfinement.
The system created in such collisions is large relative to the hadronic scale; it is characterized by energy densities
  and temperatures sufficient to generate a deconfined multiparticle state.

Instead of concentrating on the microscopic {\it mechanism} of confinement, a fruitful strategy has been to focus
  on ``soft'' (low momentum or large length scale) observables to probe the properties of the bulk {\it matter} itself.
Of particular interest is the QCD equation of state (EoS), quantifying the relationships between intrinsic quantities such
  as energy density and temperature.
The nature of these relationships depends on the phase, and as in condensed matter physics,
  much may be learned by focusing on the transitions between phases.

\begin{figure}[]
{\centerline{\includegraphics[width=0.42\textwidth]{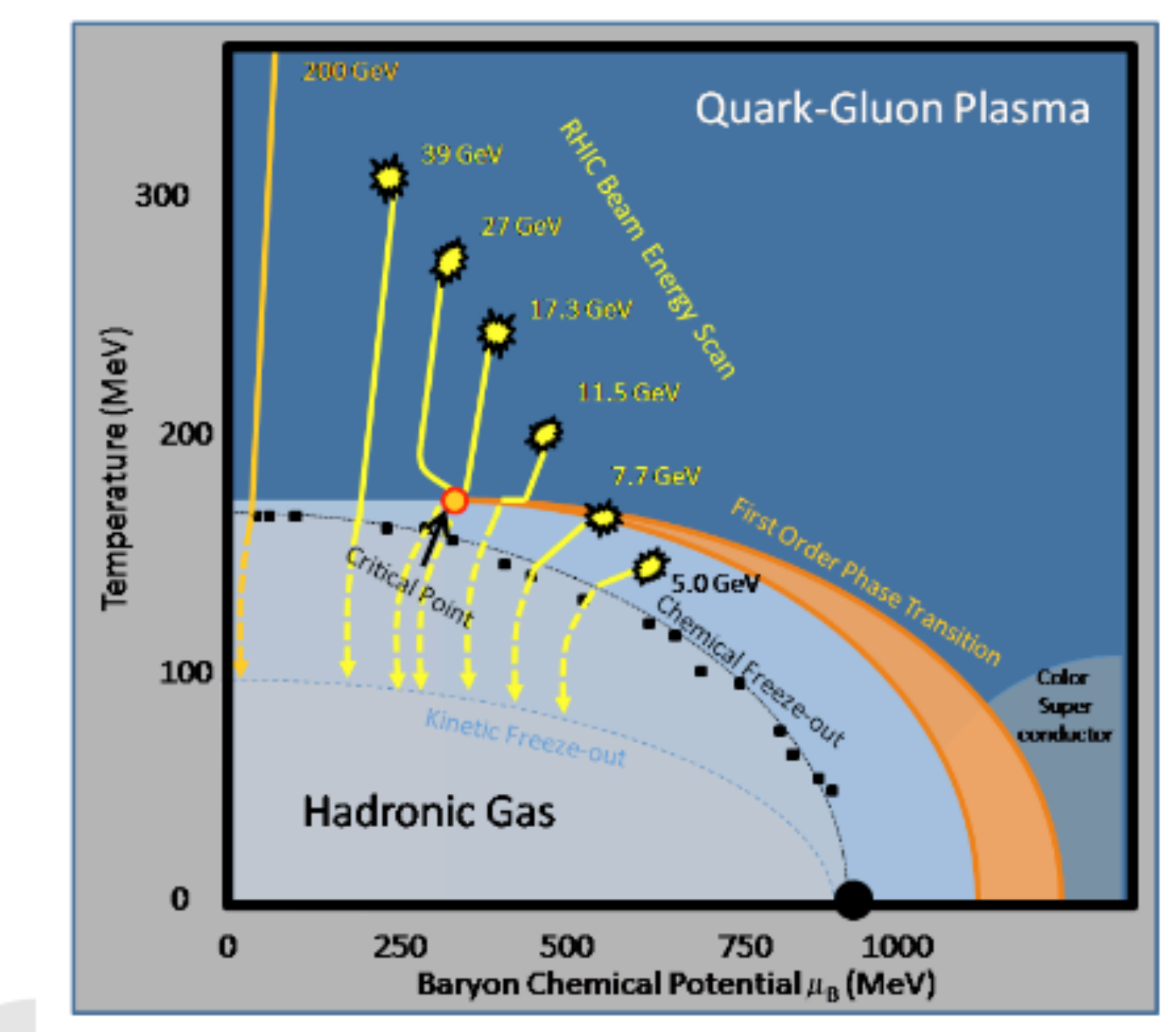}}}
\caption{Sketch of the QCD phase diagram in the $T-\mu_B$ plane.  An estimate of the critical point and first-order
  transition line is indicated, as are possible trajectories of systems created in collisions with energy $\sqrt{s_{NN}}$.
  From~\cite{Aggarwal:2010cw}.
\label{fig:PhaseDiagram}
}
\end{figure}

In nuclear collisions with center-of-mass energy $\sqrt{s_{NN}}\sim 100$~GeV, there is considerable evidence
  that the matter passes into the deconfined phase during some part of its evolution~\cite{Adams:2005dq,Adcox:2004mh,Back:2004je,Arsene:2004fa,Tannenbaum:2006ch}.
Data do not show evidence of a latent heat associated with the transition (e.g. prolonged lifetime~\cite{Pratt:1986cc,Rischke:1996em}
  or zero-pressure mixed phase), and there is general agreement that the transition is a smooth cross-over
  at $T\sim 150$~MeV and $\mu_B~30$~MeV.
Generic considerations~\cite{Stephanov:2004wx} lead to the expectation that a critical point and first-order phase transition
  will be found elsewhere-- at lower temperatures and larger chemical potentials-- on the $T-\mu_B$ plane.
Figure~\ref{fig:PhaseDiagram} shows a schematic of the QCD phase diagram~\cite{Aggarwal:2010cw}.

Locating and studying these landmarks on the landscape of the QCD phase diagram would reveal much
  about the interaction, including fundamental quantities such as latent heats, critical exponents and universality class.
Creating a system that samples these more interesting conditions requires lowering the collision energy.
For this reason, experiments at RHIC have embarked on a major program to map out the energy dependence of
  experimental observables, under fixed detector and analysis conditions~\cite{Aggarwal:2010cw}.
Ideally, this exercise parallels that of a condensed matter physicist, precisely controlling the
  temperature of a material and measuring its resistance, a precipitous drop clearly marking the transition
  to a superconductor.

However, unlike the condensed matter lab sample, the system created in RHIC collisions is highly dynamic
  and far from infinite; phase transition signals in heavy ion collisions will be more subtle.
The dynamics of any bulk system are dictated by its EoS, which encodes the non-trivial landmarks on the phase diagram
  discussed above.
Thus, it is particularly interesting to identify bulk observables sensitive to the dynamics and
  EoS.

Two-particle intensity interferometry probes the final-state geometry of a nuclear collision at
  the femtometer scale~\cite{Lisa:2005dd}.
Detailed measurements of this geometry as a function of transverse momentum ($p_T$), rapidity ($y$)
  and particle mass, have revealed a rich spacial substructure of the system due to pressure-driven
  bulk collective flow.
It has long been recognized that azimuthally-sensitive studies in non-central collisions
  generally yield more insight than do azimuthally-integrated measurements.
Unfortunately, while momentum-space measurements of azimuthal anistropies are plentiful at several collision energies,
  there are few such measurements in coordinate-space.

In this paper, we compare anisotropic shape measurements in heavy ion collisions with similar studies
  in two other strongly-interacting systems at very different spatial and thermal scales.
We discuss the physics associated with anisotropic shapes in these collisions, and how these shapes are measured.
Several theoretical transport calculations are used to show the sensitivity of the final shape of the source
  on the underlying physics, and are compared with existing measurements.
Particularly interesting is an ``anomalous'' shape measurement at about $\sqrt{s_{NN}}\approx 20$~GeV,
  an energy region where several threshold-like behaviours have been reported~\cite{Gazdzicki:2010iv}.
Given the extreme paucity of shape measurements, however, it is difficult to conclude much from the data at this point.
Rather, our work represents a call for a much more detailed shape analysis at low RHIC energies.

\section{Anisotropic shapes as a probe of three strongly-coupled systems}
\label{sec:analogs}
As evidenced by this special volume, there is increasing recognition of connections between the physics of ultra-relativistic heavy ion
  collisions and other quite different fields.
In this section, we briefly discuss three systems of vastly different scales and decidedly distinct physical constituents;
  these are listed in table~\ref{tab:Scales}.
Despite their differences, they share a striking resemblence.
In all cases, a strongly-interacting system is initially prepared in a spatially anisotropic state and then allowed to evolve.
The shape at a later time reveals important physics driving the dynamics of the matter.

\subsection{Anisotropic shape evolution in a cold atomic gas}
\label{sec:coldGas}

\begin{figure}
\begin{center}
\includegraphics[width=0.33\textwidth]{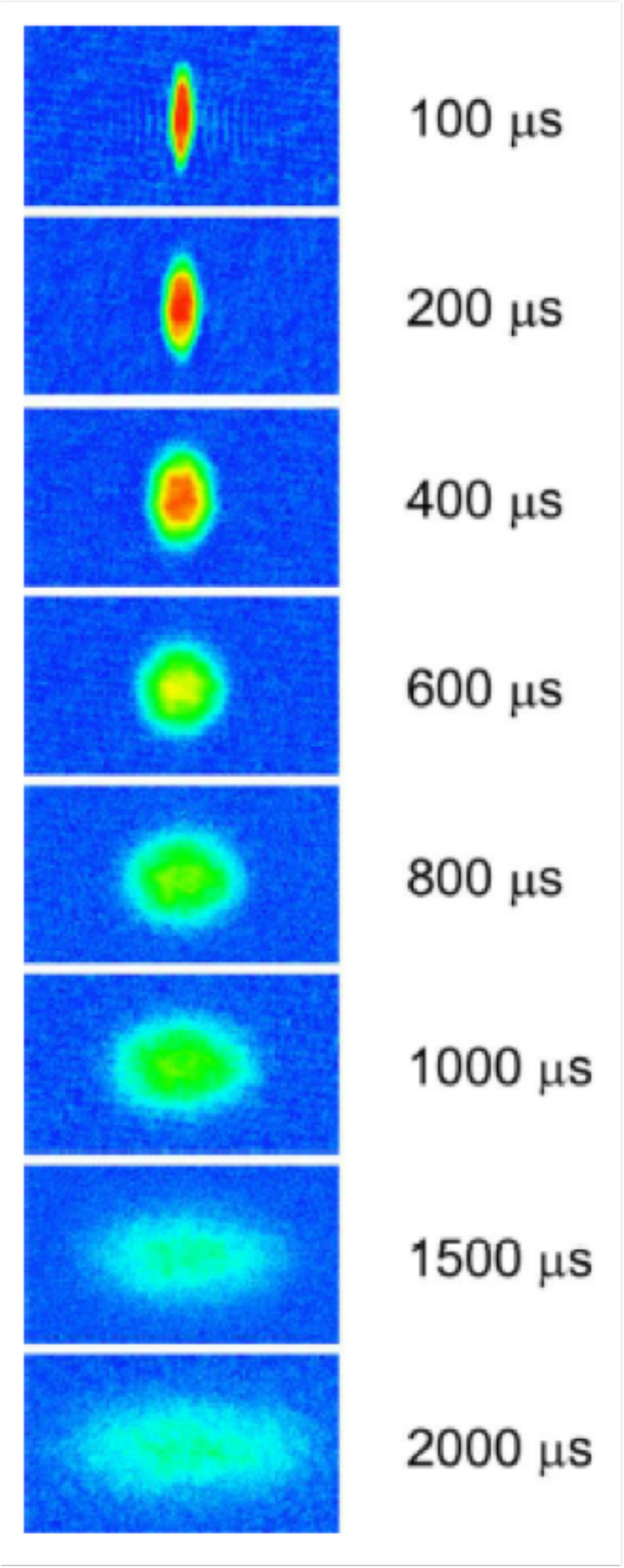}
\hspace{10mm}
\includegraphics[width=0.45\textwidth]{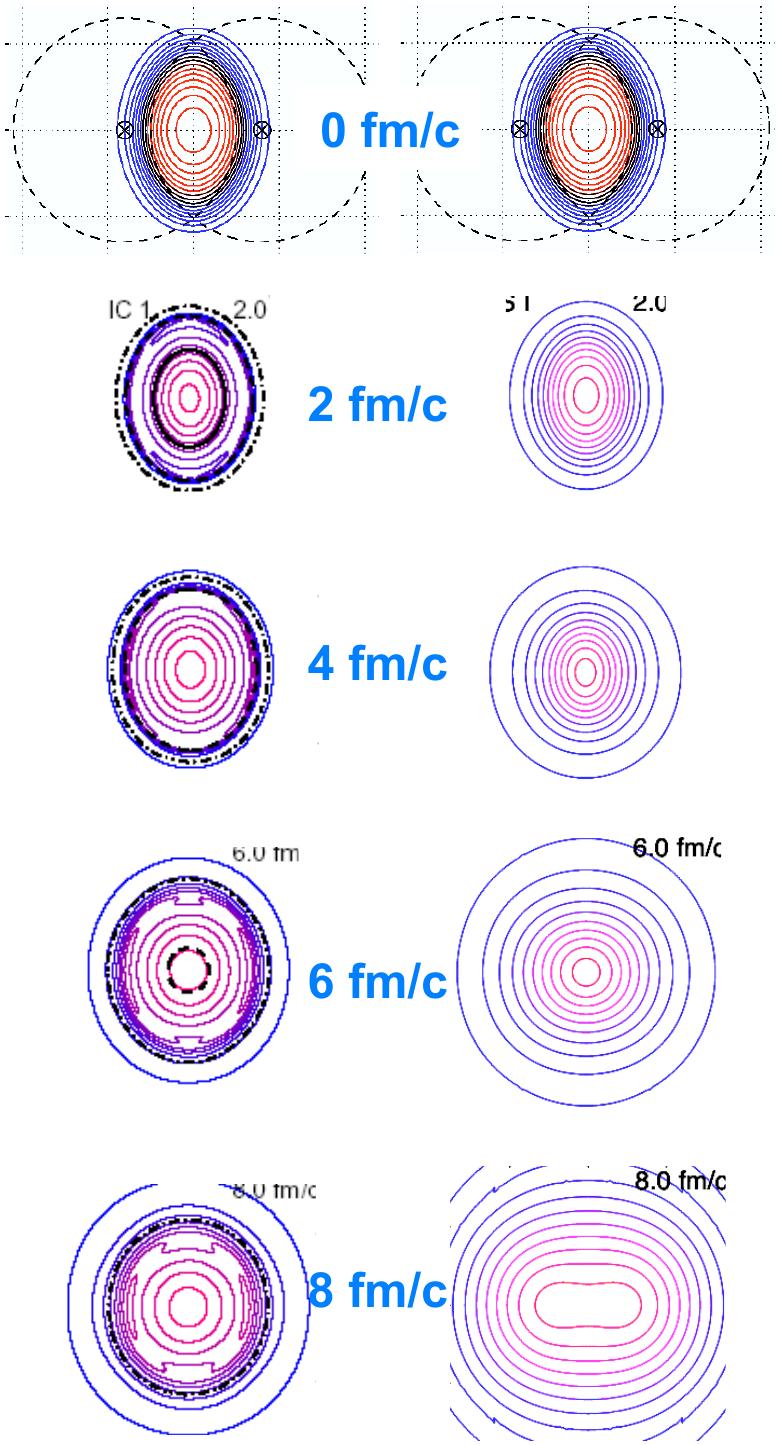}
\caption{Time evolution of spatial anisotropy two strongly-coupled systems.  {\bf Left:} A degenerate Fermi gas
  of ultra-cold Li atoms released from an anisotropic trap.  From~\cite{O'Hara:2002zz}.  Reprinted with permission from AAAS.
  {\bf Right:} Hydrodynamical calculation of the evolution of a Au+Au collision at $\sqrt{s_{NN}}=130$~GeV.
  Evolution on the right corresponds to an equation of state (EoS) for an ideal
  massless gas.  On the left, the EoS includes a first-order transition between hadronic and QGP phases.
  From~\cite{Kolb:2003dz}, reprinted with permission.
\label{fig:Ohara}
\label{fig:KolbHydroEvolution}
}
\end{center}
\end{figure}

The first connection to the bulk evolution in heavy ion collisions and that of cold atomic systems, was pointed out several years ago~\cite{Kolb:2003dz,Shuryak:2003ty}.
In particular, in measurements by O'Hara {\it et al}~\cite{O'Hara:2002zz}, a degenerate Fermi gas of ultra-cold $^6{\rm Li}$ atoms
  is held in a spatially-anisotropic magnetic trap.
The trap is removed, and the shape of the system is measured at a later time; Figure~\ref{fig:Ohara}
  shows the state of the system at different times after the trap is released.
(The measurement of the system shape actually destroys the cold gas, so the panels of fig.~\ref{fig:Ohara}
  are in reality different gas samples released from identical traps.)

For the system depicted in Figure~\ref{fig:Ohara}, a pumping laser has been used to maximize the
  inter-atom interaction cross section.
Thus, like the ultra-hot partonic system created at RHIC, the ultra-cold degenerate gas is a strongly-coupled
  quantum fluid, starting from an elongated ellipsoidal configuration and allowed to expand into the surrounding vacuum.
As with the QGP at RHIC, the higher pressure gradients along the short direction of the initial shape (horizontal in the
  figure) lead to a stronger expansion in that direction; with the passage of time, the system becomes more round and after
  some time ($\sim 600$ $\mu$s in the figure) even reverses the sense of its elongation.

The similarity between these measurements and those of ``elliptic flow'' in heavy ion collisions~\cite{Voloshin:2008dg}
  has been noted by others~\cite{Kolb:2003dz,Shuryak:2003ty}.
However, measurements of elliptic flow are restricted to anisotropies in momentum space; a more direct connection is
  made in space-time.
Eleven orders of magnitude smaller and 21 orders faster, the heavy-ion analog is shown in the hydrodynamical calculations of
  Figure~\ref{fig:KolbHydroEvolution}~\cite{Kolb:2003dz}.
Here, the initial anisotropy of the system is generated by the finite impact parameter of the collision.
As with the cold atomic gas, the system rapidly expands preferentially along its shorter axis.

\begin{table}[t!]
\begin{tabular}{|l|l|l|l|}
\hline
system                             & T (K)~~~       & length (m) & time (s) \\
\hline
cold atoms                         & $10^{-6}$   & $10^{-4}$    &$ 10^{-3}$  \\
electrical plasmas in crystals~~   & $10^{5}$    & $10^{-7}$    & $10^{-12}$  \\
heavy ion collisions               & $10^{12}$   & $10^{-15}$    & $10^{-24}$  \\
\hline
\end{tabular}
\caption{Characteristic scales for three strongly-coupled systems.
Despite their widely different scales and matter characteristics, their evolution and methods used to study them are strikingly similar.
They are prepared in an anisotropic state, expand hydrodynamically, and their final shape is studied to determine their underlying physical properties.
}
\label{tab:Scales}
\end{table}


In both systems, the strength of the expansion is driven by pressure gradients which are, in turn, determined by the energy
  density through the equation of state and thermodynamic state of the system.
The inversion of the aspect ratio in the cold gas system, seen about 700~$\mu$s after its release, signals
  a strongly interacting phase, semi-quantitatively understood as a superfluid state~\cite{O'Hara:2002zz,Menotti:2002zz}.
For the heavy ion case, the calculations on the left and right in Figure~\ref{fig:KolbHydroEvolution} begin with identical initial
  energy distributions; only the EoS is different.
If the EoS of a massless gas is assumed (right column), the pressure is large and the expansion rapid.
For an EoS featuring a first-order phase transition, the pressure gradients are initially large (in the QGP
  phase), then very small (as the system passes through the mixed phase) and finally of moderate strength (in
  the confined phase).
Since different regions of the system pass through these phases at different times, the flow pattern is complex.
What is clear is that the freeze-out shape in coordinate space is very sensitive both to the EoS of the hot matter
  of interest, and to the timescale over which its evolution takes place.
It is this shape that is extracted by azimuthally-dependent femtoscopic measurements discussed in this paper.

\begin{figure}[t]
\begin{center}
\includegraphics[width=0.45\textwidth]{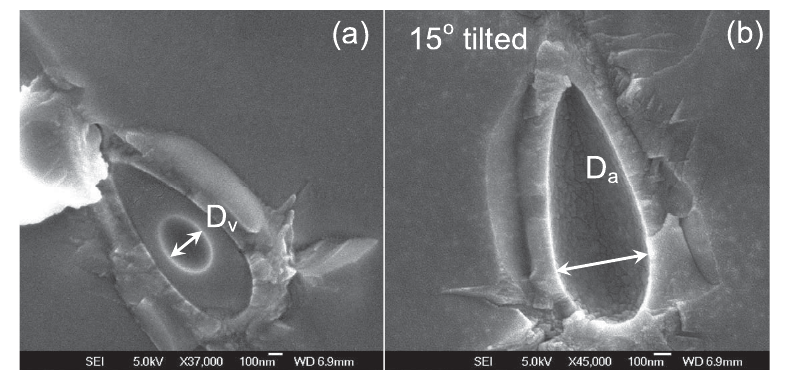}
\hspace{\fill}
\includegraphics[width=0.45\textwidth]{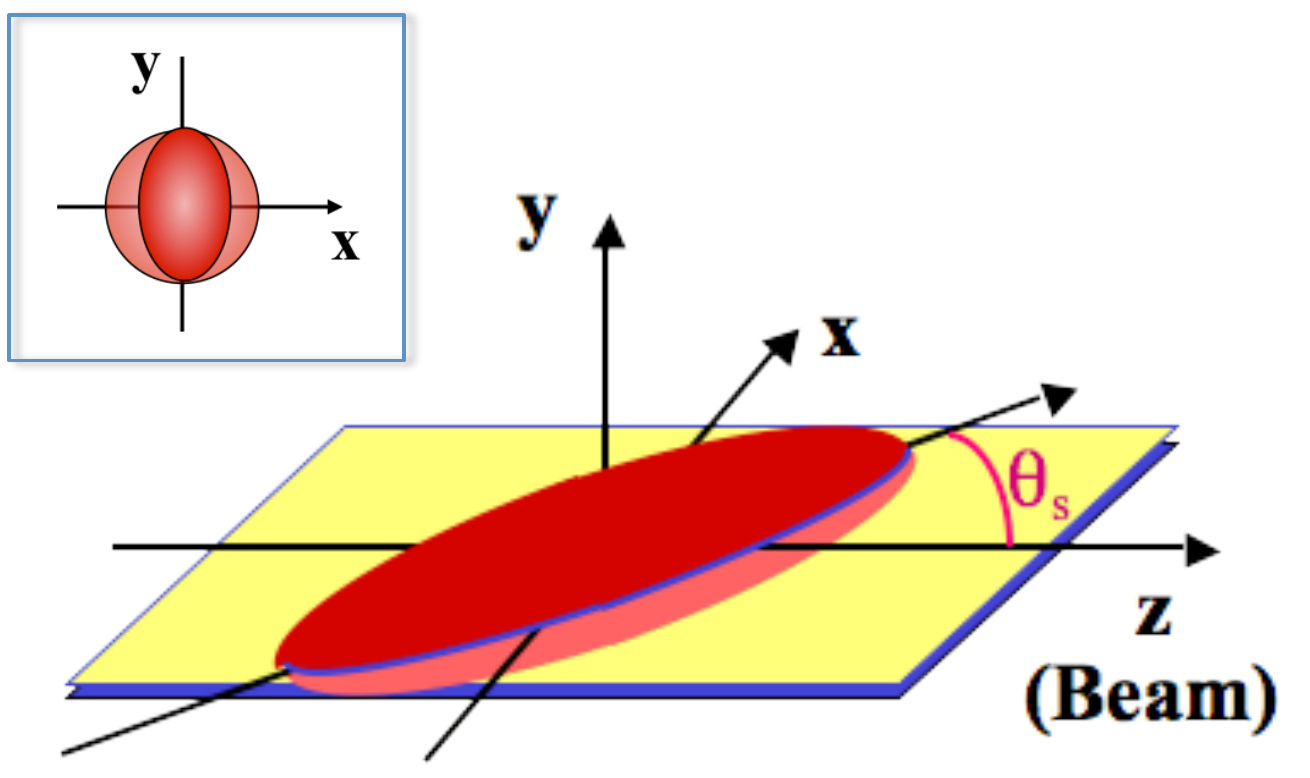}
\caption{Anisotropic final state of the evolution of dynamic plasmas.
{\bf Left:} Scanning electron microscope images of the aftermath of a microexplosion in sapphire
    produced by a 150-fs, 100-nJ laser pulse.
    Reprinted with permission from~\cite{PhysRevLett.96.166101}; copyright (2006) by the American Physical Society.
    \label{fig:microexplosion}
{\bf Right:} Simplified parameterization of the freezeout distribution in a heavy ion collision,
      as an ellipsoid tilted with respect to the beam axis.  Inset: Projection of the distribution
      in the transverse plane.  
      \label{fig:TiltedEllipseCartoon}}
\end{center}
\end{figure}

\subsection{Anisotropic shape evolution in an electrically-deconfined plasma}
\label{sec:microExplosion}

Recently, there has been much activity in the study of dynamic condensed matter systems under extreme conditions.
This work bears even greater similarity to our study of spatial anisotropies in heavy ion collisions.

The study of condensed matter systems under extreme pressure has a long history; recently, using diamond anvils,
pressures of up to 0.1~TPa ($10^6$~atm) can be achieved under static conditions in the
  laboratory, resulting in a surprising diversity of new materials~\cite{McMillan:NatureMaterials2002}.
Exploring even more extreme conditions-- 10~TPa or larger-- requires explosive generation of a transient system 
  such as are done at the National Ignition Facility~\cite{Lindl:PhysicsOfPlasmas2004} or in table-top experiments
  in which sub-ps laser pulses generate ``microexplosions'' under the surface of Sapphire crystals~\cite{PhysRevLett.96.166101} crystals or fused silica~\cite{Mermilod-Blondin:APL2009}.

The study of these microexplosions parallels strikingly the study of femtoexplosions in heavy ion
  collisions.
In the initial state, the matter is in the charge-confined (atomic) state.
Upon rapid deposition of extreme energy density ($10^{17}~\rm{J/m^3}$),
  a charge-deconfined plasma is generated within a few fs, at temperatures of $10^5-10^6$~K.
The plasma expands rapidly ($\sim$~ps), cooling as it does so, and returns to charge-confined degrees of freedom.
Plasma hydrodynamics and two-component ``blast-wave'' pictures~\cite{PhysRevB.76.024101}
  are used to describe and understand the source evolution~\cite{PhysRevLett.96.166101}.


With huge changes in physical scales and ``color charge'' replacing ``electric charge,'' the above describes the situation
  with RHIC collisions rather well, down to the blast-wave parameterizations~\cite{Retiere:2003kf}.
In both cases, too, the final-state anisotropy carries important physical information.
The anisotropic final-state geometry of a microexplosion is measured directly by a scanning electron microscope; c.f.
  Figure~\ref{fig:microexplosion}.
In a heavy ion experiment, it is the final-state momenta that are directly measured, and
  azimuthally-sensitive two-particle intensity interferometry must be used to measure the coordinate-space geometry.

Since the first proof-of-principle microexplosion experiments, there has been considerable activity to extract the equation
  of state of the matter-- the plasma state, phase transitions, etc.
The approach taken is essentially identical to the one we now propose at RHIC: to measure the final-state anisotropy as the
  initial energy of the system is varied, and compare the results to transport calculations with different EoS.
Even the cryptic names of the equations of state~\cite{Hallo:AppliedPhysA2008} (QEOS, SESAME 7387, etc) are reminiscent of those used in RHIC studies.

\subsection{Anisotropic shape evolution in hot QCD matter}
\label{sec:ShapesHI}

The case on which we shall focus henceforth is the anisotropic evolution of the hot matter generated in the overlap
  zone of two colliding heavy nuclei; this is indicated in the right panels of Figure~\ref{fig:KolbHydroEvolution}.
Here, we introduce the anisotropies of interest and the physics driving their evolution.
The situation with heavy ion collisions bears more resemblence to that of the microexplosions of section~\ref{sec:microExplosion}
  than to the cold atoms discussed in~\ref{sec:coldGas}, since the experimenter cannot freely choose the time to measure the system anisotropy.
When particles decouple from the medium created in a heavy ion collision, they are said to ``freeze out.''
Only the final state of the system-- after it has expanded and frozen out-- is available for examination; its temporal evolution must be modeled.

The anisotropy of the hot zone in a heavy ion collision has two sources.
Firstly, the beam direction ($\hat{z}$) is clearly special;
  both in momentum- and coordinate-space, the hot source is extended in the $\hat{z}$.
Collisions at finite impact parameter break the remaining symmetry in the azimuthal variable around the beam direction.
The so-called reaction plane is the plane spanned by the impact parameter (oriented in the $\hat{x}$-direction in this
  work) and the beam direction.
Figure~\ref{fig:TiltedEllipseCartoon} shows a plausible if simplistic sketch of the hot matter produced in a non-central heavy ion
  collision, containing the minimal set of possible anisotropies-- different length scales in each direction, and a tilt of
  the source away from the beam axis.

Of particular interest is the transverse eccentricity of the source, mentioned already in section~\ref{sec:coldGas}.
This eccentricity may be quantified by $\epsilon \equiv \left(\sigma_y^2-\sigma_x^2\right)/\left(\sigma_y^2+\sigma_x^2\right)$,
  where $\sigma_{x,y}$ are characteristic scales of the system in and out of the reaction plane, respectively,
  and will be discussed in more detail shortly.
As discussed there, and seen in Figure~\ref{fig:KolbHydroEvolution}, the final state eccentricity is determined
  by both the anisotropic pressure gradient and the system lifetime; increasing either or both of these results
  in a lower (possibly negative) $\epsilon$.

The other major feature of the freezeout distribution is the tilt of its major axis, relative to the beam direction.
Such tilts are ubiquitously produced in three-dimensional simulations of heavy ion collisions.
At low energies ($\sqrt{s_{NN}}\approx 4$~GeV), $\theta_s\approx30^\circ$~\cite{Lisa:2000xj}; its sign
  discriminated between competing explanations of momentum-space anisotropies for charged pions~\cite{Lisa:2000ip}.

\begin{figure}[t!]
\begin{center}
\includegraphics[width=0.48\textwidth]{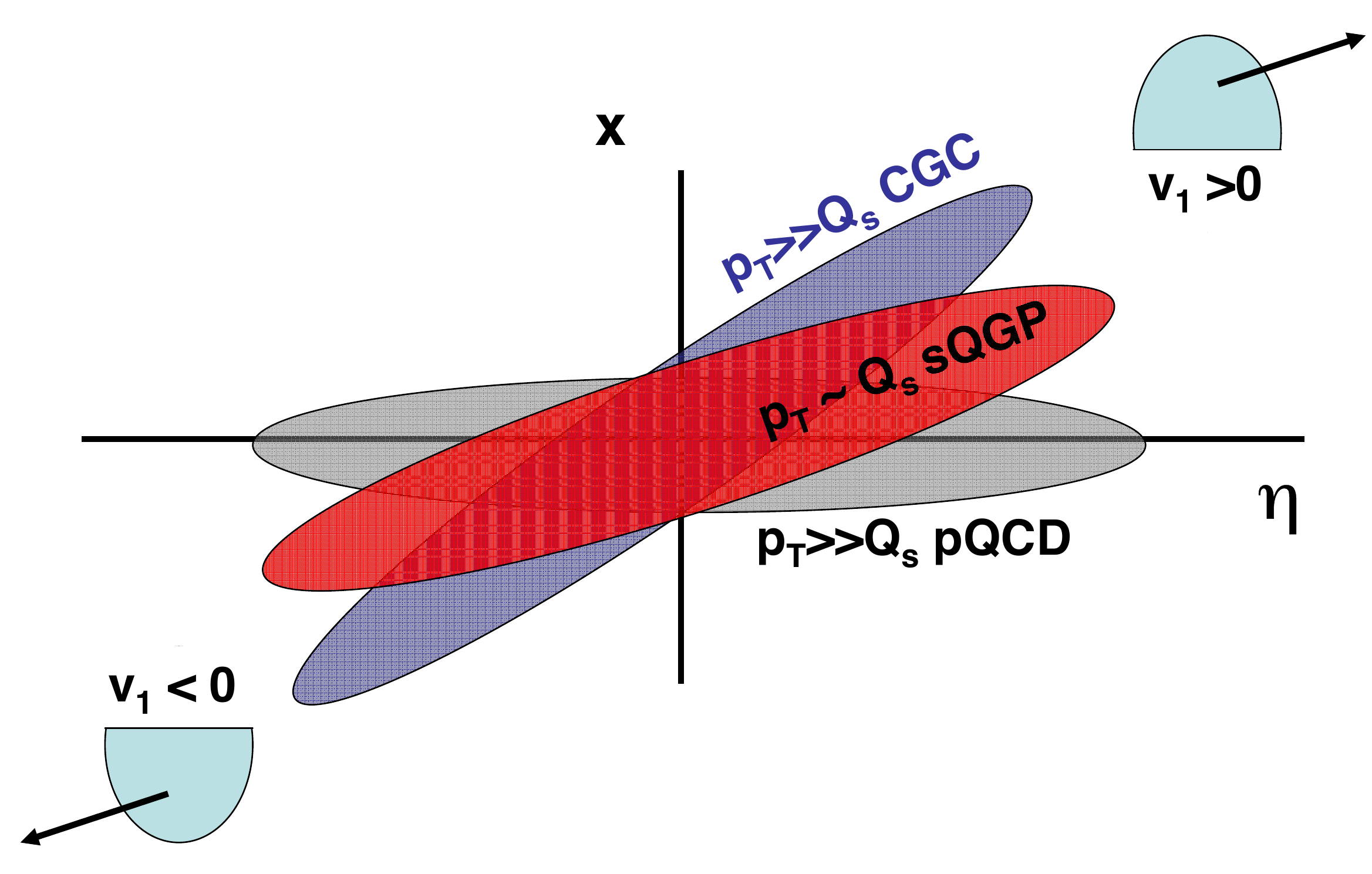}
\includegraphics[width=0.48\textwidth]{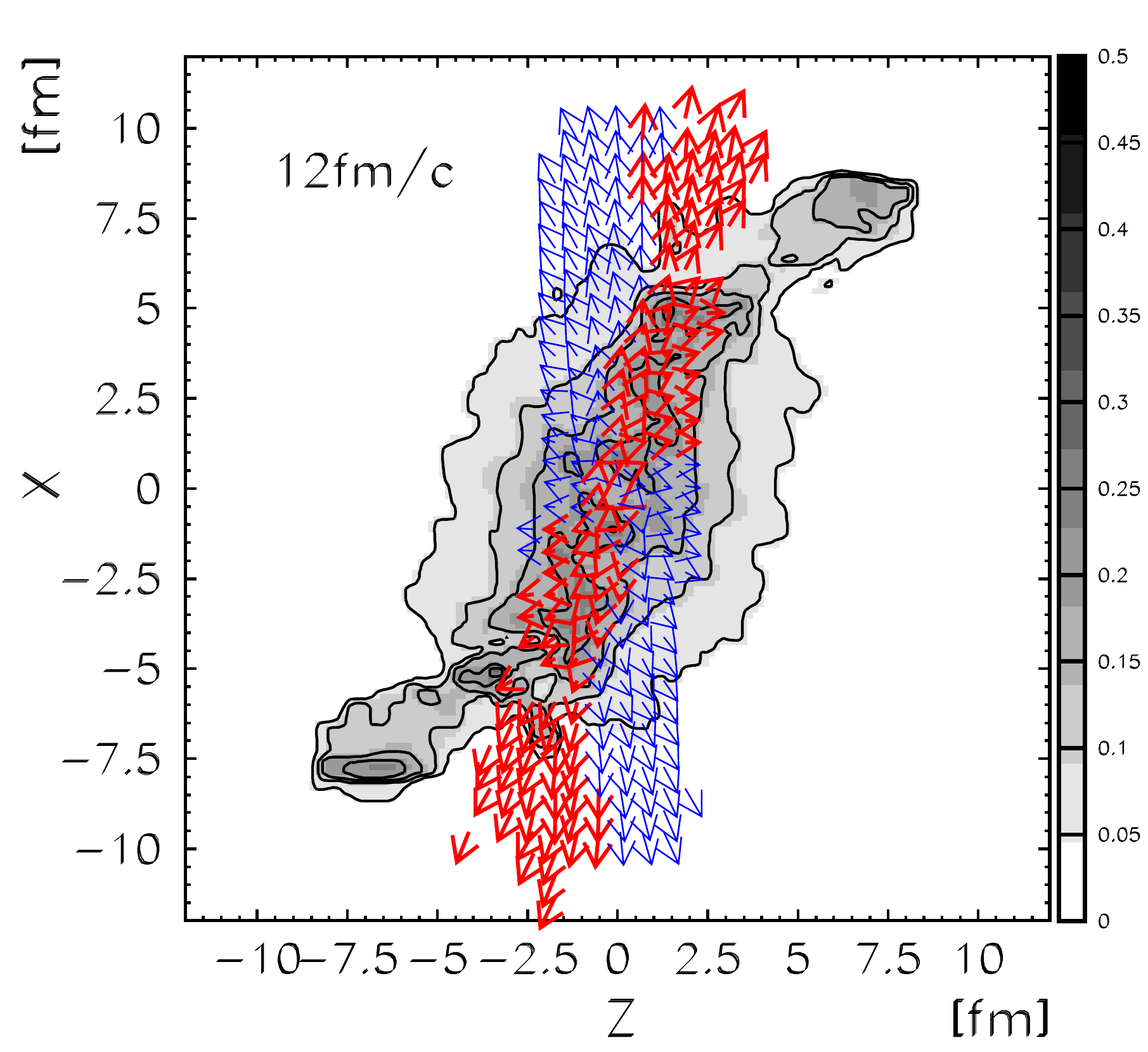}
\caption{Important aspects of a tilted QGP.  {\bf Left:} The tilt of the bulk ``twisted sQGP'' in general will not coincide
  with that of the hard partonic collisions.  The interplay between the two can distinguish different initial-state scenarios
  and strongly affects jet quenching signals.  Reprinted with permission from~\cite{Adil:2005bb}; copyright (2006) by the American Physical Society.
  {\bf Right:} The ``anti-'' (or third) flow component (blue arrows) arises from the expansion of the tilted source along its short axis
   and partially cancels the normal flow (red arrows).  Reprinted with permission from~\cite{Brachmann:1999xt}; copyright (2000) by the American Physical Society.
\label{fig:TwistAnti}
}
\end{center}
\end{figure}

No tilt meaurements have yet been made at ultra-relativistic energies, but several theorists have pointed to its importance.
In what they have termed the ``twisted sQGP,'' Adil and collaborators~\cite{Adil:2005bb,Adil:2005qn}  
  emphasize the importance of the tilt of the strongly-coupled quark-gluon plasma created in non-central collisions; c.f. Figure~\ref{fig:TwistAnti}.
In particular, the source of hard partonic scatterings (leading to jets) will have a different tilt than that
  of the plasma itself.
The interplay between the two tilts is crucial to obtain fully three-dimensional jet tomography~\cite{Adil:2005qn}.
The hard-partonic tilt depends on the initial-state model, while the tilt (or ``twist'')
  of the sQGP can be measured by techniques discussed in the next section.

The tilt of the sQGP leads to important signals in the bulk sector, as well.
In three-dimensional fluid dynamic calculations, a tilted hot zone generates a collective structure known as the
  third flow~\cite{Csernai:1999nf,Magas:2000cm}  or anti-flow~\cite{Brachmann:1999xt}, as it expands preferentially along
  its shortened axis; c.f. Figure~\ref{fig:TwistAnti}.
This tilt or ``torque'' can arise naturally in a wounded nucleon initial condition~\cite{Bozek:2010bi,Bozek:2010vz}.
This third flow component combines nontrivially with ``normal'' directed flow, leading to partial cancelation at low
  energies~\cite{Brachmann:1999xt} as suggested in the right panel of Figure~\ref{fig:TwistAnti}, while the third flow dominates at the LHC~\cite{Csernai:2011gg}.
Directed, or first-order flow signals in momentum space~\cite{Voloshin:2008dg} are among the most important bulk
  phenomena, sensitive to the earliest, densest stages of the collision~\cite{Snellings:1999bt} which may or may not
  be thermalized.
Spatial tilt measurements-- even if only of the freezeout distribution-- will need to be combined with momentum-only
  analyses to disentangle the dynamics of this crucial stage of the collision.

Generic expectations for the collision energy dependence of freezeout shapes seem straightforward.
The eccentricity, $\epsilon$, is affected by pressure and timescale.
One expects both the lifetime and the energy density of the system to increase with increasing $\sqrt{s_{NN}}$.
Thus-- if the relationship between pressure and energy density (the EoS) remains fixed-- it is natural to expect
   $\epsilon$ to decrease monotonically with $\sqrt{s_{NN}}$.
The tilt is a manifestly non-boost-invariant aspect of the QGP created in the collision.
Directed flow measurements at all energies confirm that the dynamics of heavy ion collisions are never, strictly speaking, boost invariant.
Even the hope that the system is ``essentially'' boost-invariant at midrapidity may be easily shattered if a finite tilt angle is measured there.
Nevertheless, due to the increased elongation of dynamics along the beam direction, it is natural to expect a monotonic decrease of $\theta_s$
  with $\sqrt{s_{NN}}$ as well.

\section{Measuring source anisotropy in heavy ion collisions}

In the two parallel cases discussed in section~\ref{sec:analogs}, the final spatial source anisotropy is measured directly.
Of course, the spatial and temporal scales involved in a heavy ion collision render such measurements impossible.
Instead, spatial sizes and shapes are extracted via two-particle femtoscopy~\cite{Lisa:2005dd}.
This technique exploits the connection between the measured two-particle relative momentum correlation function $C(\vec{q})$
  and the spatial separation distribution $S(\vec{r})$, according to the Koonin-Pratt equation~\cite{Koonin:1977fh},
\begin{equation}
\label{eq:KP}
C(\vec{q}) = \int d^3r^\prime S(\vec{r}^\prime) |\phi(\vec{q}^\prime,\vec{r}^\prime)|^2  ,
\end{equation}
where $\phi(\vec{q},\vec{r})$ is the two-particle wavefunction as a function of the relative momentum and separation
  in the pair center of mass system, $\vec{q}\equiv\vec{p}_1-\vec{p}_2$
  $\vec{r}^\prime\equiv\vec{x}_1^\prime-\vec{x}_2^\prime$.
Femtoscopy has been used extensively to map the space-time structure of heavy ion collisions for a quarter century;
  details of the method and the physics learned have been discussed elsewhere, e.g.~\cite{Lisa:2005dd}.

Significantly complicating any femtoscopic analysis is the fact that, in Equation~\ref{eq:KP}, $C(\vec{q})$ and $S(\vec{r})$
  may-- and in reality always do-- depend on the pair momentum $\vec{K}\equiv\frac{1}{2}\left(\vec{p}_1+\vec{p}_2\right)$.
This is a consequence of space-momentum correlations that arise from collective flow and means that particles
  emitted with a given velocity measure only part of the source.
Reconstructing the ``whole'' source from the various fragments is highly non-trivial and ultimately model-dependent.
For the moment, we ignore this fact and assume that the pion emission probability from any space-time point of the source
  is independent of its momentum.
We return to the issue of position-momentum correlations at the end of the section.

The most common femtoscopic analysis in heavy ion collisions correlates identical pions.
Often, the separation distribution is assumed Gaussian; in this case, ignoring Coulomb and other
  complications~\cite{Lisa:2005dd}, the correlation function itself is Gaussian,
  and fitted with
\begin{equation}
C\left(\vec{q}\right) = 1+\lambda\exp\left(-\sum_{i,j=o,s,l}q_iq_jR^2_{i,j}\right) .
\label{eq:GaussianFit}
\end{equation}
The indices indicate the components of the relative momentum vector in the 
  Bertsch-Pratt ``out-side-long'' coordinate
  system, in which the ``long'' direction is parallel to the colliding beams, ``out'' is parallel to the pair transverse
  momentum $\vec{K}_T\equiv\frac{1}{2}(\vec{p}_{T,1}+\vec{p}_{T,2})$, and ``side'' is perpendicular to ``out'' and ``long.''
Use of this coordinate system is motivated by the fact that spatial and temporal aspects of the source are more easily disentangled~\cite{Pratt:1986cc,Bertsch:1989vn}.
The fit parameters $R^2_{i,j}$ are squared ``HBT radii'' that characterize the three-dimensional size and shape of the separation
  distribution.

\begin{figure}
\begin{center}
\includegraphics[width=0.48\textwidth]{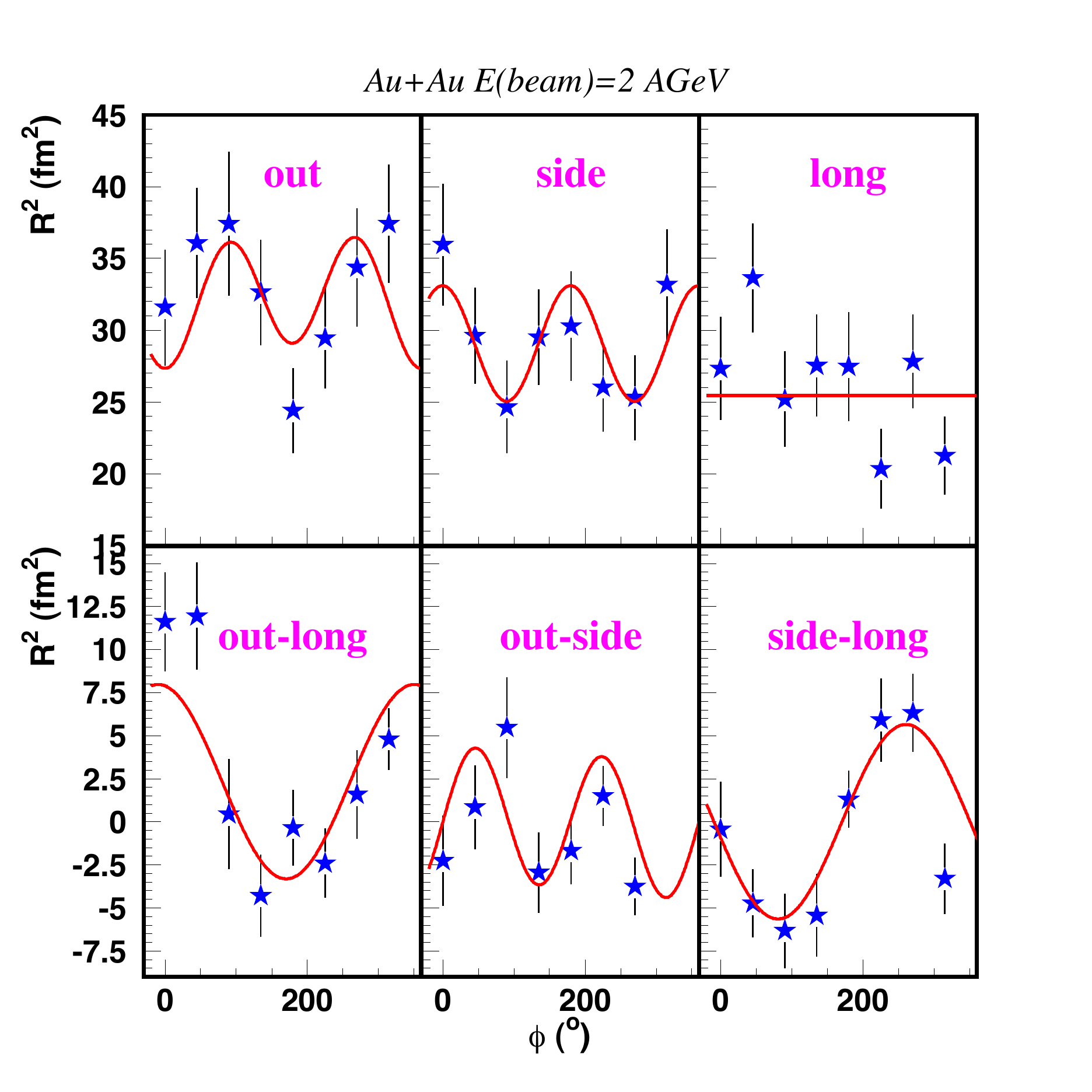}
\includegraphics[width=0.48\textwidth]{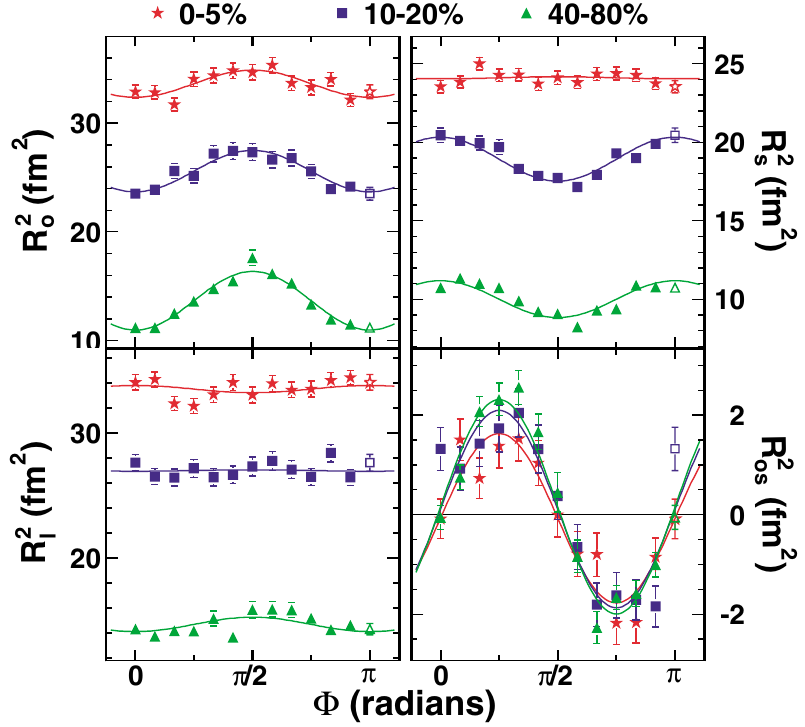}
\caption{Oscillating HBT radii at the lowest and highest measured energies.
  {\bf Left:} $\sqrt{s_{NN}}=2.35$~GeV Au+Au collisions with impact parameter $b\approx 5$~fm.~\cite{Lisa:2000xj}
  {\bf Right:} $\sqrt{s_{NN}}=200$~GeV Au+Au collisions of varying centrality.
   Reprinted with permission from~\cite{Adams:2003ra}; copyright (2004) by the American Physical Society.
\label{fig:RadiiData}
}
\end{center}
\end{figure}

In an azimuthally sensitive femtoscopic study, these HBT radii are measured as a function of the pair angle
  $\phi_p\equiv\angle\left(\vec{K}_T,\vec{b}\right)$.
Since the ``out'' and ``side'' directions rotated relative to the $x$ and $y$ directions by $\phi_p$, even
  in the simplest case of Figure~\ref{fig:TiltedEllipseCartoon}, one expects oscillations in the $R^2_{i,j}\left(\phi_p\right)$.
Indeed, such oscillations in non-central collisions have been clearly observed; Figure~\ref{fig:RadiiData}
  shows HBT radii measured at the lowest ($\sqrt{s_{NN}}=2.2$~GeV)~\cite{Lisa:2000xj} and highest ($\sqrt{s_{NN}}=200$~GeV)~\cite{Adams:2003ra} energy collisions explored by
  these analyses.
Suppression of oscillations due to the finite reaction-plane resolution may be corrected for~\cite{Heinz:2002au}.
Similar to the analysis of momentum flow, Fourier moments of the oscillating radii are extracted.
\begin{equation}
\label{eq:HBTradii-FCs}
R^2_{\mu,n} =
\begin{cases}
\langle R^2_\mu(\phi_p) \cos(n\phi_p) \rangle  (\mu = o, s, l, ol) \\
\langle R^2_\mu(\phi_p) \sin(n\phi_p) \rangle  (\mu = os, sl)
\end{cases}.
\end{equation}

Very few femtoscopic studies to date have been performed relative to the event-wise reaction plane; instead,
  the direction of the impact parameter is ignored, and the resulting length scales represent an azimuthal
  average of all collisions.
In this case, the pair emission angle $\phi_p$  is meaningless, and no oscillations are measured.
Furthermore, the ``cross-term'' radii $R^2_{i,j\neq i}$ vanish by symmetry~\cite{Heinz:2002au} in these analyses.

A Gaussian functional form is the most commonly-used parameterization of the pion-emitting source in heavy ion collisions;
  it is characterized by three scales: a lifetime, and spatial scales in the longitudinal and transverse directions.
The minimal generalization in the azimuthally-sensitive case is a Gaussian ellipsoid with three principle axis lengths $\sigma_{x,y,z}$,
  a timescale $\sigma_t$, and a tilt angle $\theta_s$ relative to the beam direction.
\begin{equation}
\fl
f\left(x,y,z,t\right) \sim \exp\left( -\frac{\left(x\cos\theta_s-z\sin\theta_s\right)^2}{2\sigma_{x^\prime}^2} \right.
                -\frac{y^2}{2\sigma_y^2}   
                \left. -\frac{\left(x\sin\theta_s+z\cos\theta_s\right)^2}{2\sigma_{z^\prime}^2}
                -\frac{t^2}{2\sigma_t^2}\right) .
\label{eq:GaussianSource}
\end{equation}
The primes on $\sigma_{x^\prime}$ and $\sigma_{x^\prime}$ denote that these are the lengths of the primary axes of the ellipse,
  the source widths in the tilted coordinate system (c.f. Figure~\ref{fig:TiltedEllipseCartoon}).

In the simplest case in which Equation~\ref{eq:GaussianSource} describes the pion-emitting source, its anisotropy parameters
  are simply related to the measured HBT radii.
The spatial eccentricity along the beam axis is given by~\cite{Retiere:2003kf}
\begin{equation}
\epsilon \equiv \frac{\sigma_y^2-\sigma_x^2}{\sigma_y^2+\sigma_x^2} = 2\cdot\frac{{R}^2_{s,2}}{R^2_{s,0}} .
\label{eq:ExtractingEpsilon}
\end{equation}
Here, $\sigma_{x}$ (without the prime) denotes the width of the source in the $x$-direction, as seen along the beam
  direction, not along the (tilted) major axis of the ellipse.

If the first-order oscillations in $R^2_{sl}$ is measured,
  the tilt angle is estimated by combining several Fourier coefficients~\cite{Lisa:2000ip,Mount:2010ey},
\begin{equation}
\theta_s = \textstyle{1\over 2} \tan^{-1}\left(\frac{-4{R}^2_{sl,1}}{R^2_{l,0}-R^2_{s,0}+2{R}^2_{s,2}}\right) ,
\label{eq:ExtractingTheta}
\end{equation}
and the transverse eccentricity in the ``natural'' frame tilted relative to the beam axis is~\cite{Mount:2010ey}
\begin{equation}
\fl
 \epsilon^\prime \equiv \frac{\sigma_y^2-\sigma_{x^\prime}^2}{\sigma_y^2+\sigma_{x^\prime}^2} 
 = \frac{2{R}^2_{s,2}\left(1+\cos^2\theta_s\right)+\left(R^2_{s,0}-R^2_{l,0}\right)\sin^2\theta_s-2{R}^2_{sl,1}\sin2\theta_s}
                       {R^2_{s,0}\left(1+\cos^2\theta_s\right)+\left(2{R}^2_{s,2}+R^2_{l,0}\right)\sin^2\theta_s+2{R}^2_{sl,1}\sin2\theta_s}
\label{eq:ExtractingEpsilonPrime}
\end{equation}


As suggested in the inset to Figure~\ref{fig:TiltedEllipseCartoon}, $\sigma_{x^\prime} < \sigma_{x}$ for a simple tilted elongated 
  ellipsoid.
Hence, if the final-state emitting source retains its initial out-of-plane extension ($\sigma_y > \sigma_{x^\prime}$), as
  indicated by measurements so far, the eccentricity measured about the beam axis will be smaller than that measured about the
  tilted axis: $\epsilon < \epsilon^\prime$.

At low energies, the direction of the impact parameter $\vec{b}$ can be estimated easily, thanks to a relatively strong
  first-order anisotropy in momentum-space-- the so called ``directed flow;'' in this case, $\phi_p$ has a meaningful range $[0,2\pi]$.
At RHIC energies, on the other hand, the first-order momentum-space anisotropy is weak, while the second-order momentum-space anisotropy
  (``elliptic flow'') is much easier to measure.
Thus, at the higher energies, only the plane that contains $\vec{b}$ is defined, but not the direction of $\vec{b}$ itself;
  this corresponds, in Figure~\ref{fig:TiltedEllipseCartoon}, to identifying the yellow reaction plane, but not distinguishing
  $\pm\hat{x}$.
In this case, $\phi_p$ is measured only modulo $\pi$ and first order oscillations like $R^2_{sl,1}$ cannot be measured.
Spatial information on the source tilt $\theta_s$ and eccentricity in the source's natural coordinate system $\epsilon^\prime$
  are inaccessible in such analyses.
Similarly, two-dimensional transport calculations such as 2+1-dimensional boost-invariant hydrodynamics~\cite{Kolb:2003dz} are implicitly
  blind to any tilt structure.

As mentioned at the beginning of this section, strong position-momentum correlations, due to collective flow or other sources,
  imply that pion pairs measured at a given $\phi_p$ do not sample the entire pion-emitting zone, but only a selected ``homogeneity
  region''~\cite{Akkelin:1995gh}.
In {\it principle,} the correspondence between the homogeneity regions and the ``whole'' source can be almost arbitrary, so that
  extracting the shape of the latter through measurement of the former is neccessarily model dependent.
However, studies with reasonable blast-wave parameterizations~\cite{Retiere:2003kf} and realistic transport calculations~\cite{Mount:2010ey}--
  both of which feature strong flow and non-trivial correspondence between homogeneity regions and the entire source--
  indicate that Equations~\ref{eq:ExtractingEpsilon}-\ref{eq:ExtractingEpsilonPrime} are good to a model-dependent systematic uncertainty of $\sim 30\%$.


\section{Compilation of experimental results}
\label{sec:data}

\begin{table}[b!]
\begin{tabular}{|l|l|l|l|}
\hline
Experiment                                                & $\sqrt{s_{NN}}$ (GeV)    &  centrality (\%)                         & rapidity                     \\
\hline
AGS/E895~\cite{Lisa:2000xj}                               & 2.35, 3.04, 3.61         & (7.4-29.7)                               & $|y|<0.6$                    \\
\hline
\multirow{2}{*}{SPS/CERES~\cite{Adamova:2008hs}}          & \multirow{2}{*}{17.3}    & (7.5-10)$\oplus$(10-15)$\oplus$(15-25)   & \multirow{2}{*}{$-1<y<-0.5$} \\
                                                          &                          & and (10-15)$\oplus$(15-25)                  &                           \\
\hline
\multirow{2}{*}{RHIC/STAR~\cite{Adams:2003ra}}            & \multirow{2}{*}{200}     & (5-10)$\oplus$(10-20)$\oplus$(20-30)     &  $|y|<0.5$                   \\
                                                          &                          & and (10-20)$\oplus$(20-30)               &                              \\
\hline
\end{tabular}
\caption{Measurements of the anisotropic shapes from heavy ion collisions.
   The third column indicates which centrality bins were averaged, to obtain the shape parameters of Figures~\ref{fig:ThetaExcitationFunction} and~\ref{fig:EpsilonExcitationFunction}.
   See text for details.
\label{tab:experiments}}
\end{table}

Three experiments, listed in Table~\ref{tab:experiments}, have published azimuthally-sensitive pion HBT radii.
All estimated the impact parameter of the collision based on charged particle multiplicity.
Since system anisotropy clearly depends on the impact parameter, it is important to
  compare collisions with similar centrality.
In order to best compare results from the $7.4-29.7\%$ centrality cut of E895 (corresponding to $b=4-8$~fm),
  several centrality cuts were combined, for the higher energy measurements.

It is worthwhile to describe in detail how centrality bins were merged,
  since the comparison between shapes from different collision energies is important for our message.
The most relevant centrality bins reported by the STAR Collaboration are for $5-10\%$, $10-20\%$ and $20-30\%$ of total cross section.
We combine data from these three bins as
\begin{equation}
\label{eq:combine}
\epsilon^{\rm STAR}_{(5-10)\oplus(10-20)\oplus(20-30)} \equiv \frac{1\times\epsilon^{\rm STAR}_{(5-10)}+2\times\epsilon^{\rm STAR}_{(10-20)}+2\times\epsilon^{\rm STAR}_{(20-30)}}{1+2+2} .
\end{equation}
Here, the weighting factors (1,2,2) account for the fact that the $5-10\%$ bin contains half the number of events as either of the other two
  bins listed.
This selection includes more central events-- in particular, events in the $5-7.4\%$ centrality range-- than are included in the E895 cuts.
Therefore, the value $\epsilon^{\rm STAR}_{(5-10)\oplus(10-20)\oplus(20-30)}=0.081\pm0.006$ should be considered a lower bound on the shape compared
  with E895.
An upper bound may be obtained by combining only the two more peripheral bins: $\epsilon^{\rm STAR}_{(10-20)\oplus(20-30)}=0.094\pm0.007$.

The relevant centrality ranges reported by CERES are $7.5-10\%$, $10-15\%$ and $15-25\%$ of the total cross section.
A fair range to use in our comparison is between $\epsilon^{\rm CERES}_{(7.5-10)\oplus(10-15)\oplus(15-25)} = 0.035\pm0.018$
  and $\epsilon^{\rm CERES}_{\oplus(10-15)\oplus(15-25)} = 0.043\pm0.020$.

\begin{figure}
\begin{center}
\includegraphics[width=0.6\textwidth]{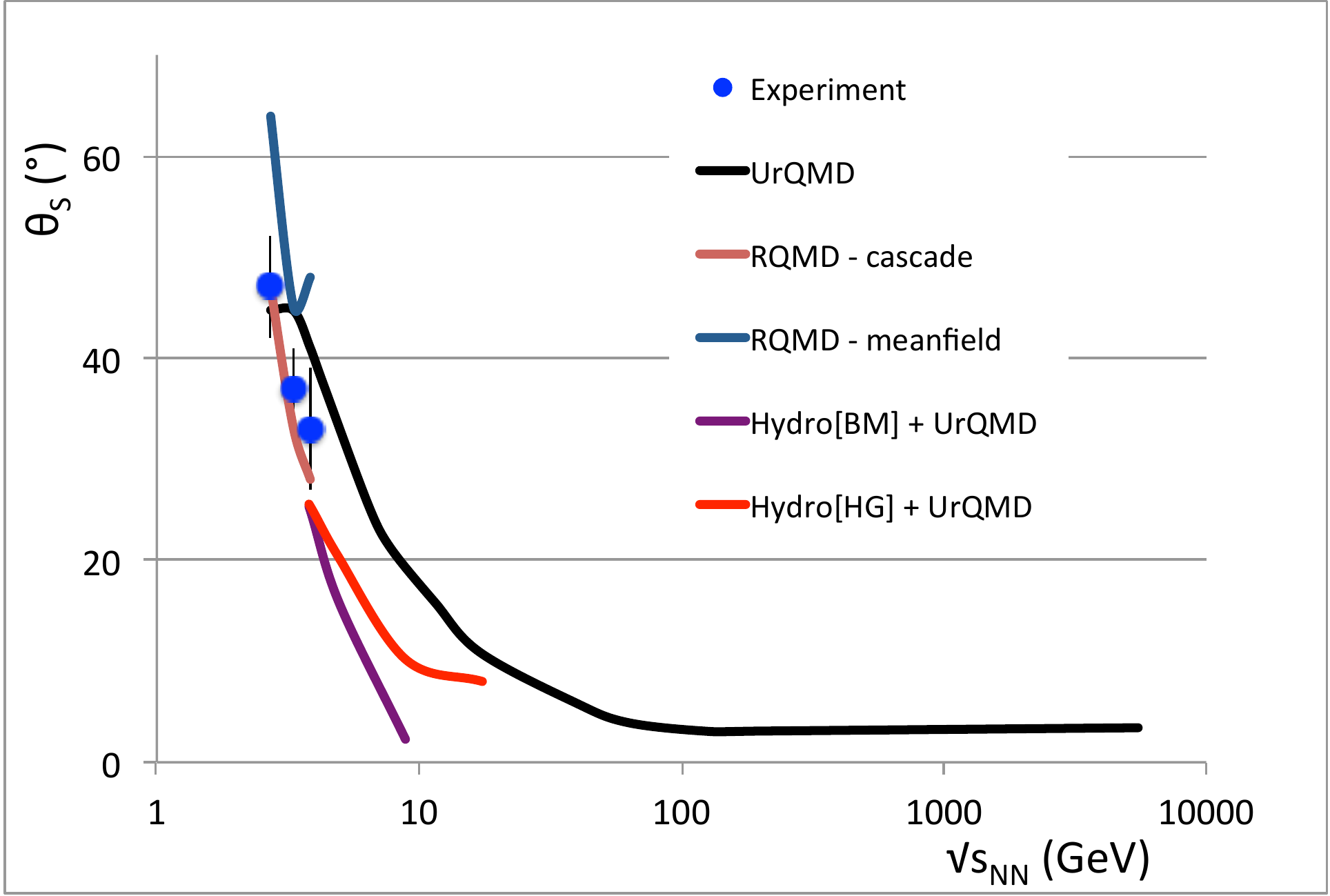}
\caption{Source tilt relative to the beam direction vs. energy for midcentral heavy ion collisions.
See sections~\ref{sec:data} and~\ref{sec:models} for a discussion of the experimental data
  and model calculations, respectively.
\label{fig:ThetaExcitationFunction}
}
\end{center}
\end{figure}

All measurements focused on low-momentum pions ($p_T\approx0.25$~GeV/c), for which the formulae~\ref{eq:ExtractingEpsilon}-\ref{eq:ExtractingEpsilonPrime}
  work best~\cite{Retiere:2003kf}.
STAR and E895 measurements center on midrapidity, where participant contributions should be maximal, while the CERES
  measurement is somewhat backwards in the center of mass frame.
Since HBT measurements typically vary slowly with rapidity, this difference is unlikely to affect CERES' shape estimation,
  but a measurement at midrapidity would provide a better comparison with the other experiments.

E895 measured HBT radii relative to the first-order event plane (i.e., the direction of the impact parameter); results
  for $\sqrt{s_{NN}}=2.35$~GeV are shown on the left panel of Figure~\ref{fig:RadiiData}.
The spatial tilt is shown in Figure~\ref{fig:ThetaExcitationFunction}.
The tilt is strikingly large at these low energies and drops with energy, consistent with the expectation~\cite{Bjorken:1982qr}
  that collisions become increasingly boost-invariant (at least near midrapidity) with increasing energy.
It will be important to extend tilt measurements to higher energies, since a finite $\theta_s$ is manifestly ``boost-variant,''
  even {\it at} $y=0$.
If $\theta_s$ is more than a few degrees, boost-invariant models may not be valid and would at least require double-checking
  with true three-dimensional calculations.

\begin{figure}
\begin{center}
\includegraphics[width=0.75\textwidth]{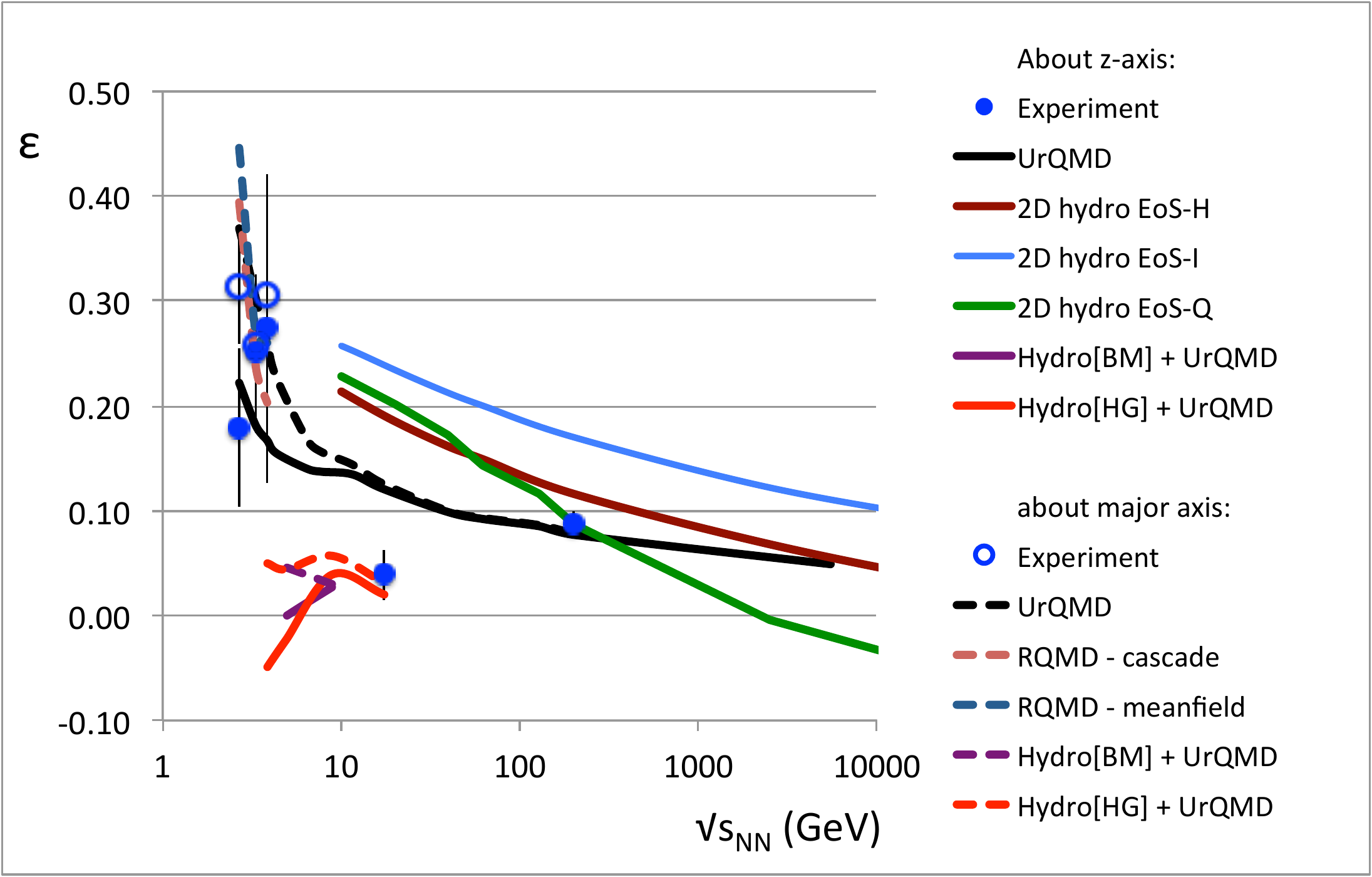}
\caption{Source eccentricity vs. energy for midcentral heavy ion collisions.
See sections~\ref{sec:data} and~\ref{sec:models} for a discussion of the experimental data
  and model calculations, respectively.
\label{fig:EpsilonExcitationFunction}
}
\end{center}
\end{figure}

Figure~\ref{fig:EpsilonExcitationFunction} shows the measurements from the experiments listed in Table~\ref{tab:experiments}.
Filled symbols indicate $\epsilon$, the eccentricity relative to the beam axis (c.f. Eq.~\ref{eq:ExtractingEpsilon}), while open symbols
  indicate the eccentricity in the natural frame of the source (Eq.~\ref{eq:ExtractingEpsilonPrime}), measured only by E895.
For the CERES and STAR datapoints, the average of the upper and lower bounds discussed above are plotted, with the difference between the bounds and the statistical
  errorbars added, to be conservative.
The non-monotonic behaviour of $\epsilon\left(\sqrt{s_{NN}}\right)$ is intriguing.
As discussed in Section~\ref{sec:ShapesHI}, rather general considerations lead to the expectation of a monotonic decrease of $\epsilon$
  with energy.
The unexpected dip in Figure~\ref{fig:EpsilonExcitationFunction} occurs in the energy region in which phase transition ``threshold'' effects have
  been reported~\cite{Gazdzicki:2010iv} and around which some speculate that heavy ion collisions sample the non-trivial features sketched in 
  Figure~\ref{fig:PhaseDiagram}; c.f~\cite{Aggarwal:2010cw}.

To contribute our own speculation, we note that such non-monotonic behaviour could arise from one of two effects, both related to 
  a first-order phase transition.
Firstly, an extended lifetime due to the transition would allow the system to evolve further towards a round shape
  (c.f. Figure~\ref{fig:Ohara}), causing a dip just around the threshold energy.
Using this simplistic scenario to explain the data, the CERES datapoint at $\sqrt{s_{NN}}\approx 17$~GeV lies near the threshold energy.
Alternatively, we may fix the lifetime and consider effects of the stiffness of the EoS, quantified by the speed of sound in the medium, $c^2_s=\frac{\partial P}{\partial e}$,
  where $P$ and $e$ are the pressure and energy density, respectively.
At low energies, the system is in the hadronic phase, and $c_s^2\sim\frac{1}{6}$; as $\sqrt{s_{NN}}$ increases,
  pressure gradients increase in proportion to the energy deposited in the system; here, $\epsilon$ would fall with $\sqrt{s_{NN}}$.
Near the threshold energy, the system may spend much of its time in the mixed phase, for which $c^2_s=0$; here, the system shape would evolve little from its
  initial, large value.
As the energy increased still further, the system spends most of its time in the deconfined plasma phase, for which $c_s^2\sim 1/3$, and $\epsilon\left(\sqrt{s_{NN}}\right)$
  again falls monotonically.
Using this second simplistic scenario to explain the data, the STAR datapoint at $\sqrt{s_{NN}}= 200$~GeV lies just above the threshold energy.

However, such simplistic considerations only serve to stimulate more sophisticated treatment with theoretical tools.
We consider such tools in the next section.

\section{Transport model calculations}
\label{sec:models}
Transport models are commonly used to simulate the dynamics of a heavy ion collision.
They fall into two broad categories-- hydrodynamical calculations and Boltzmann transport models.

As the name implies, hydrodyanmic calculations model the nuclear matter in terms of one or more fluids, each fluid
  cell characterized by thermodynamically intensive quantities such as temperature, chemical potential, and pressure.
Particles {\it per se} play no role in the dynamics; rather, they appear at the end of the simulation, as the
  fluid cells produce final-state particles according to an approximate prescription~\cite{Cooper:1974mv}.
A major advantage of hydrodynamic calculations is that the equation of state
  must be explicitly defined.
The relationships between bulk quantities-- pressure, energy density, temperature, baryon number, etc--
  close the set of hydrodynamic equations, which are otherwise ``trivial'' continuity and conservation laws.
These relationships characterize and quantify the state of matter under different conditions and are
  the mathematical implementation of phase diagrams such as that in Figure~\ref{fig:PhaseDiagram}.

While hydro calculations focus explicitly on bulk quantities, microscopic Boltzmann transport calculations take the opposite
  tack:  the creation, scatterings and demise of each particle is followed.
Such calculations offer several advantages over hydrodynamic ones: they are manifestly three-dimensional (important in the present context); assumptions of thermalization
  and equilibrium, justified or not, are not required; event-by-event fluctuations-- potentially large in such finite systems and of increasing interest in the field-- are
  implicitly included; bulk and shear viscosity are automatically included in the calculation; finally, the final
  state of the system-- freeze-out-- occurs ``naturally'' on a particle-by-particle basis, without recourse to arbitrary criteria (e.g. a fixed local temperature) setting
  the freeze-out hypersurface.
The chief drawback of Boltzmann transport calculations is their complexity; the dynamics is influenced by the decay and reaction details of 
  several thousand hadron species, as well as the string phenomenology needed to describe non-hadronic degrees of freedom
  in the ultra-high density limit at which $2\rightarrow2$ scattering simulations break down.


We use several transport calculations to simulate the evolution of a heavy-ion collision from its initial
  out-of-plane-deformed shape, to the final shape probed by femtoscopy.

\subsection{Two-dimensional boost-invariant hydrodynamics}
\label{sec:2dHydro}

Among the earliest and most exciting successes in the RHIC program was the degree to which ideal hydrodynamic calculations predicted the magnitude
  and details (mass- and $p_T$-dependence) of bulk collective behavior in non-central heavy ion collisions~\cite{Kolb:2003dz,Voloshin:2008dg,Heinz:2009xj}.
Here, we use one of the most successful models-- AZHYDRO~\cite{Kolb:2000sd}, a (2+1)-dimensional ideal hydrodynamic model.
This model uses a common simplifying assumption~\cite{Shuryak:1980tp,Bjorken:1982qr} that the initial conditions and subsequent dynamics are boost-invariant; all
  interesting physics takes place in the transverse plane.
For our purposes, this means that the model assumes a non-tilted source ($\theta_s=0$), and only $\epsilon$ is calculable.


It is by now rather clear that boost-invariant ideal hydrodynamic calculations are invalid at energies below about $\sqrt{s_{NN}}\approx 20$~GeV~\cite{Kolb:2003dz}
  as (i) the densities achieved are inconsistent with a zero-mean-free-path approximation, and (ii) the system is not boost-invariant.
Nevertheless, we will perform calculations over a wide range of energies, including very low ones, to map the excitation function and explore the
  EoS dependence of the shape parameters.

In order to extend the model into both the LHC high energy regime as well as 
low energy heavy-ion collisions, the initialization routine must be tuned to 
the appropriate collision energies. The initial temperature distribution was
parameterized through the initial transverse entropy profile geometrically 
using an optical Glauber calculation discussed in~\cite{Kolb:2003dz}.

Within the optical Glauber model calculation, the density of wounded nucleons 
($N_{w}$) and binary nucleon-nucleon  collisions ($N_{\rm bin}$) are estimated in 
the transverse plane.  The total entropy density is then a superposition of the 
``soft'' wounded nucleon density and the ``hard'' binary collision density, 
appropriately scaled \cite{Kolb:2001qz,Kharzeev:2000ph} to match both the 
charged hadron multiplicity and centrality dependence observed in experiments 
(25\% hard contribution).  The multiplicity is simply the momentum integrated
 particle distribution at a given rapidity $y$ (typically at midrapidity, 
$y=0$) \cite{Kolb:2003dz,Kolb:2001qz}. Furthermore, the for consistency, as 
the initial entropy density (and consequently the initial temperature) changes 
for each collision, the thermalization time $\tau_0$ is correspondingly 
changed in order to keep the ``uncertainty relationship'' 
($\tau_0T_0\approx 1$) constant \cite{Kolb:2003dz,Frodermann:2007ab}. The
multiplicity is then matched to the expected charged multiplicity of each 
collision energy.

To generate the observable momentum distribution of particles, a freeze-out 
criterion is needed at the end of the hydrodynamic simulation. The Cooper-Frye 
formalism \cite{Cooper:1974mv} postulates a sudden transition from a perfect 
local equilibrium to free-streaming for all strongly interacting particles in 
a particular fluid cell at a given kinetic freeze-out condition.  While this 
can be described via a dynamic comparison of scattering rates, here we use a 
constant energy density $e = 0.075$ GeV/fm$^3$ which corresponds to a 
constant temperature isotherm which varies depending on the equation of state 
used to describe the matter.  For all equations of state used in these 
simulations, the freeze-out temperature is $T_f\approx $~130 MeV.

In order to study a range of possible phases of matter, three distinct 
equations of state were used in the hydrodynamic simulations.  These
are discussed in~\cite{Kolb:2000sd} and
represent a purely hadronic state of matter 
(EOS H), a pure quark/gluon gas (EOS I), and a Maxwellian-constructed EOS Q 
which contains a first order phase transition between EOS I and EOS H. In 
\cite{Kolb:2000sd}, EOS Q was used exclusively to describe the matter created 
at $\sqrt{s_{NN}} = 130$~GeV.


HBT radii were extracted by fitting projections of two-pion correlation functions, calculated
  according to the method of~\cite{Frodermann:2006sp}.
These radii were then used in Equation~\ref{eq:ExtractingEpsilon} to extract the final-state eccentricities
  shown in Figure~\ref{fig:EpsilonExcitationFunction}.
For each EoS used, the eccentricity monotonically decreases as a function of energy, an effect both of
  increased system lifetime and pressure as $\sqrt{s_{NN}}$ increases, as discussed in section~\ref{sec:ShapesHI}.

There is considerable sensitivity to the EoS used in the calculation.
Using the initialization procedure discussed above, use of the stiff equation of state, EoS-I with $c_s^2=\frac{1}{3}$, results in a much
  more out-of-plane shape-- i.e. one that has not evolved much from the initial overlap shape.
The shape evolves considerably more when using the softer EoS-H ($c_s^2=\frac{1}{6}$).
Since, for a given energy density pressure gradients are proportional to $c_s$, these results suggest that effects of system
  lifetime dominate over those of pressure, in these calculations.
This conclusion is consistent with the results when using EoS-Q.
These shapes track closely with those of EoS-H for low $\sqrt{s_{NN}}$ where the system is dominated by the
  hadronic phase.
At around $\sqrt{s_{NN}}$, the threshold effect of the ``soft'' mixed phase ($c_s^2=0$) become apparent, increasing the system lifetime
  and further decreasing $\epsilon$.

It is clear from Figure~\ref{fig:EpsilonExcitationFunction} that the phase transition is needed to explain the single RHIC datapoint-- at
least in this model.
Extension of the shape excitation function to higher (LHC)  energies will be important to further
  constrain the EoS.
However, many of the most important features of the EoS may be manifest at lower energies, such as those currently being
  explored in the RHIC energy scan.
The validity of (2+1)-dimensional models breaks down at these energies, limiting their utility in
  constraining the EoS with low-energy data.

\subsection{Microscopic Boltzmann transport}
\label{sec:URQMD}

In this work, we use two related Bolzmann transport codes.
The first is the Ultra-Relativistic Quantum Molecular Dynamics Model
  (UrQMD 3.3)\cite{Bass:1998ca,Bleicher:1999xi},
  which includes details of physical processes relevant over a huge range of energies.
For this reason, it is particularly attractive for the present study.
UrQMD is a covariant transport approach to simulate the interactions between hadrons
  and nuclei up to relativistic energies.
It is based on the propagation of nucleons and mesons accompanied by string degrees of freedom with
  interaction probabilities according to measured and calculated cross
  sections for the elementary reactions.
Hard scatterings with large momentum transfer are treated via the PYTHIA model~\cite{Sjostrand:2006za}.
For detailed comparisons of this version to
  experimental data, the reader is referred to~\cite{Petersen:2008kb}.

An earlier incarnation of this model, Relativistic Quantum Molecular Dynamics (RQMD)~\cite{Sorge:1995dp}
  was widely used over the more limited energy range of the AGS and SPS-- $\sqrt{s_{NN}}\sim 3-20$~GeV.
It is satisfying to see good consistency between the older and newer versions of the model, in terms of
  predicted shapes.

For the UrQMD calculations, $\epsilon$, $\epsilon^\prime$ and $\theta_s$ were extracted by directly fitting
  the freezeout distribution with the functional form of Equation~\ref{eq:GaussianSource}.
For the older RQMD calculations, the model output was processed through Equation~\ref{eq:KP} to generate
  correlation functions, which were fitted with Equation~\ref{eq:GaussianFit} to extract HBT radii.
These radii were then used to calculate shape parameters according to Equations~\ref{eq:ExtractingEpsilon}-\ref{eq:ExtractingEpsilonPrime}.
For these models, the shape parameters extracted using these two methods should be consistent to $\sim30\%$~\cite{Mount:2010ey}, 

The simulations reproduce the very large tilt angles measured at low $\sqrt{s_{NN}}$ and predict a sharp fall-off with energy.
The RQMD model features the possibility to include the effects of a medium-induced mean field on the trajectories of
  the hadrons during the collision; $\theta_s$ is significantly sensitive to this mean-field effect.
In particular, the spatial tilt measurements are best described when effects of the mean field are ignored (``cascade mode'').
This is interesting in light of the fact that reproducing the {\it momentum}-space tilt (``directed flow'' or $v_1$~\cite{Voloshin:2008dg})
  demands inclusion of mean-field effects~\cite{Liu:2000am}.
While measurement of spatial shapes already constrain the EoS of hot nuclear matter, combining both coordinate- and momentum-space
  shapes place even stricter constraints on the dynamics.

The final-state eccentricity, plotted in Figure~\ref{fig:EpsilonExcitationFunction}, reproduces the large $\epsilon$ (and $\epsilon^\prime$)
  values measured at the AGS, with little dependence on the nuclear mean field.
At SPS energies ($\sqrt{s_{NN}}\approx17$~GeV), tilt angles on the order of $10^\circ$ are predicted by UrQMD.
This is just about the point at which the effect of the tilt on the measured eccentricity (c.f. inset of
  Figure~\ref{fig:TiltedEllipseCartoon}) vanishes; i.e. $\epsilon\approx\epsilon^\prime$ for $\sqrt{s_{NN}}>17$~GeV in this model.
The monotonic decrease predicted by the model is not nearly strong enough to reproduce the CERES measurement, but falls rather
  smoothly to closely approach the shape measured at top RHIC energy.
At still higher energies (e.g. LHC), the model predicts a continued out-of-plane final eccentricity with little $\sqrt{s_{NN}}$-dependence.

\subsection{Hybrid hydrodynamic-Boltzmann calculation}
\label{sec:Hybrid}

Combined microscopic+macroscopic approaches are among the most
successful ideas for the modeling of the bulk properties of HICs
\cite{Andrade:2006yh,Hirano:2005xf,Nonaka:2006yn}. 
The approach that we are using here has recently been developed
and is based on the UrQMD hadronic transport approach including a
(3+1)-dimensional 
one fluid ideal hydrodynamic evolution
\cite{Rischke:1995ir,Rischke:1995mt} for the hot and dense stage of the reaction \cite{Petersen:2008dd,Petersen:2009vx}.
To mimic
experimental conditions as realistically as possible the initial
conditions and the final hadronic freeze-out are calculated using
the UrQMD approach. 
The non-equilibrium dynamics in the
very early stage of the collision and the final state interactions
are properly taken into account on an event-by-event-basis. Furthermore, the
hybrid model allows for a dynamical coupling
between hydrodynamics and transport calculation in such a way that
one can compare calculations with various EoS during the
hydrodynamic evolution and with the pure cascade calculations within
the same framework.

The coupling between the UrQMD initial state and the hydrodynamical
evolution proceeds when the two Lorentz-contracted nuclei have
passed through each other, $t_{\rm start} = {2R}/{\sqrt{\gamma^2
-1}}$ \cite{Steinheimer:2007iy}. After that, a full (3+1)
dimensional ideal hydrodynamic evolution is performed using the
SHASTA algorithm \cite{Rischke:1995ir,Rischke:1995mt}. Taking into account all
three spatial dimensions explicitly in the evolution is important to be able to study the angular
dependence of HBT radii, since effects like the longitudinal tilt of the event
plane can only be consistently considered in this way \cite{Mount:2010ey}.

Serving as an input for the hydrodynamical calculation the EoS
strongly influences the dynamics of an expanding system. Two
different equations of state are used to exemplify the differences on the
extracted HBT radii due to this
external input. One is a hadron gas equation of state (HG) with the same degrees
of freedom as in the UrQMD approach \cite{Zschiesche:2002zr}. The other one is a
bag model equation of state (BM) including a strong first order phase transition
to
the quark gluon plasma with a large latent heat \cite{Rischke:1995mt}. To see if
fluctuations in the initial state affect the result differently for different
expansion dynamics during the hydrodynamic evolution these two extreme cases
have been
chosen.

The transition from the hydrodynamic evolution to the transport approach when
the matter
is diluted in the late stage is treated as a gradual transition on an
approximated iso-eigentime hyper-surface (see \cite{Li:2008qm,Petersen:2009mz} for details).
The final rescatterings and resonance decays are taken into account in the
hadronic cascade.

As with the UrQMD calculations discussed in section~\ref{sec:URQMD}, shape parameters were extracted
  from a direct fit of the freezeout distribution with Equation~\ref{eq:GaussianSource}.
Large tilts are again predicted at low collision energies, with significant sensitivity to the EoS used in the
  calculation.
The effect of the first-order phase transition (Hydro[BM]) is clear: while at $\sqrt{s_{NN}}$ even the earliest
  dense phase of the collision is below threshold to be affected by the phase transition.
However, at larger energies, the mixed phase reduces the sideward pressure very early in the system evolution,
  reducing $\theta_s$.
It would be very interesting indeed to measure tilt angles at SPS energies.

That shapes from Hydro[HG]+UrQMD are not identical to those from ``pure'' UrQMD is at first puzzling, given that
  the EoS in the hydrodynamic phase is that used in the Boltzmann model.
This is likely a technical issue-- particles that would be emitted early in the ``pure'' UrQMD simulation are generally
  absorbed into the hydro phase in the hybrid model, only to reappear at the iso-eigentime hyper-surface mentioned above.
The details of this discrepancy are still under investigation.
Thus, for now it is best not to compare hybrid calculations to pure Boltzmann simulations, but
  to compare one hybrid calculation to the other, concluding that the EoS sensitively affects the final-state
  shape of a heavy ion collision.

Due to this increased system lifetime, the system eccentricities in the hybrid calculations are allowed to evolve to much lower values
  than those predicted by UrQMD.
Indeed, the system in its natural rotated frame is essentially round transversely ($\epsilon^\prime\approx0$), so the large tilt can even produce
  $\epsilon<0$-- an in-plane extended source, measured about the beam axis; $\epsilon$ then grows with $\sqrt{s_{NN}}$ because $\theta_s$
  decreases.
The eccentricity at SPS energy essentially reproduces the CERES result, probably an artifact of the extended lifetime effect
  discussed above.

\section{Discussion and Summary}

Locating and studying non-trivial structures (phase transitions, critical points, etc) in the phase diagram
  offers keen insights on the material under consideration.
The equation of state encodes these structures into dynamical relationships that determine a system's response
  and evolution from an externally imposed initial state.
A major program is currently underway at the RHIC facility, to vary the collision energy of heavy ion collisions
  with an eye for non-trivial energy dependence of bulk observables that might signal the presence of structures on the phase diagram.
Several sensitive observables have been proposed and are under study~\cite{Aggarwal:2010cw}.
We have discussed one such observable here-- the final-state anisotropic shape of the system in coordinate-space.

The study of such shapes is strikingly similar to similar studies of strongly-coupled systems at vastly different scales.
The parallel between the evolution of a non-central heavy ion collision and that of a strongly-coupled gas of cold atoms
  released from an anisotropic trap, has been noted before, though more in relation to mometum-space, rather than coordinate-space, shapes.
However, the heavy ion situation bears much more resemblance to the study of ``micro-explosions'' induced by femtosecond laser pulses
  on crystals.
In both cases, a charged-confined system is, on very short timescales, raised to an energy density sufficiently high to generate
  a charge-deconfined plasma.
The plasma responds hydrodynamically-- expanding and cooling until the system returns to its original, charged-confined phase.
The equation of state is extracted by comparing the anisotropic shape of the final state with transport calculations with different EoS.
The energy scan program at RHIC is following the lead of studies of these micro-explosions, varying the initial energy of the system
  as a sensitive way to probe the EoS.

Whereas the shapes of cold atomic gases or micro-explosions can be measured directly, two-particle intensity interferometry
  is the most direct probe of the space-time structure of evolving matter on the femtometer scale.
We have discussed how system shapes are obtained from such measurements and the very few measurements that have been made to date.

The shape excitation function, sparse though it is, features an unexpected minimum at $\sqrt{s_{NN}}=17$~GeV, an energy around which other
  ``anomalous'' behaviour has been reported~\cite{Gazdzicki:2010iv}.
Based on qualitative arguments, we speculated that a first-order phase transition might cause the minimum.
This minimum is not predicted by any transport model, even those that include a first-order phase transition.
It is good to keep in mind, however, that none of the models we considered are perfect-- the Boltzmann models
  do not reproduce the large flow seen at RHIC, and the two-dimensional hydrodynamic models miss one of the
  prime anisotropies-- the source tilt, which is non-vanishing even at SPS energy.
In all cases, the sensitivity of the final shape of the collision to the equation of state driving
  the system's evolution, was clear.

A careful and systematic set of shape measurements at different energies is clearly warranted.
While some models (e.g. UrQMD and 2D hydro with EoS-Q) predict very similar shapes at RHIC (both close to the
  measured shape), their predictions for higher energies diverge strongly.
At lower energies, excitation functions of tilts and ellipticities form an important part of the energy
  scan program at RHIC, to search for structures on the fundamental phase diagram of QCD.

\section*{References}

\bibliographystyle{annrev}
\bibliography{citations}

\end{document}